# Temperature induced pore fluid pressurization in geomaterials

Siavash Ghabezloo[*], Jean Sulem

*Université Paris-Est, UR Navier, CERMES, Ecole des Ponts ParisTech, Marne la Vallée, France*

Keywords: thermal pressurization, pore pressure, undrained heating, drainage system, dead volume, triaxial cell, thermo-poro-elasticity

## Abstract

The theoretical basis of the thermal response of the fluid-saturated porous materials in undrained condition is presented. It has been demonstrated that the thermal pressurization phenomenon is controlled by the discrepancy between the thermal expansion of the pore fluid and of the solid phase, the stress-dependency of the compressibility and the non-elastic volume changes of the porous material. For evaluation of the undrained thermo-poro-elastic properties of saturated porous materials in conventional triaxial cells, it is important to take into account the effect of the dead volume of the drainage system. A simple correction method is presented to correct the measured pore pressure change and also the measured volumetric strain during an undrained heating test. It is shown that the porosity of the tested material, its drained compressibility and the ratio of the volume of the drainage system to the one of the tested sample, are the key parameters which influence the most the error induced on the measurements by the drainage system. An example of the experimental evaluation of undrained thermoelastic parameters is presented for an undrained heating test performed on a fluid-saturated granular rock.

[*]Corresponding Author: Siavash Ghabezloo, CERMES, Ecole des Ponts ParisTech, 6-8 avenue Blaise Pascal, Cité Descartes, 77455 Champs-sur-Marne, Marne la Vallée cedex 2, France
Email: siavash.ghabezloo@enpc.fr

# 1. Introduction

It is common in geomechanics to find situations in which an increase of temperature in a fluid saturated ground induces pore pressure increase. Considering the effect of transport of heat and fluid and the duration of event studied, two extreme cases arise, one corresponding to very large transport terms leading to a quasi drained and isothermal condition and one corresponding to very small transport terms leading to a quasi undrained and adiabatic condition. From basic mass balance laws the evolution of pore pressure is classically expressed in terms of a production/diffusion equation where appears the rate of pressure change due to temperature change under constant pore fluid mass fraction (i.e. undrained condition).

Temperature increase in saturated porous materials under undrained condition leads to thermal pressurization of the pore fluid because of the discrepancy between the thermal expansion coefficients of the pore fluid and of the pore volume. This increase in the pore fluid pressure induces a reduction of the effective mean stress, and can lead to shear failure or hydraulic fracturing. The thermal pressurization phenomenon is important in studies of rapid landslides when frictional heating tends to increase the pore pressure and to decrease the effective compressive stress and the shearing resistance of the material. The potential contribution of this phenomenon in the catastrophic loss of strength that occurred in Vaiont slide is studied by several authors (e.g. Vardoulakis, 2002, Veveakis *et al.*, 2007). The thermal pressurization phenomenon is also important in petroleum engineering where the reservoir rock and the well cement lining undergo sudden temperature changes. This is for example the case when extracting heavy oils by steam injection methods where steam is injected into the reservoir to heat the oil to a temperature at which it flows. This rapid increase of temperature could damage cement sheath integrity of wells and lead to loss of zonal isolation. This phenomenon is also important in environmental engineering for radioactive waste disposal in deep clay geological formations (Gens *et al.* 2007, François *et al.* 2009). The onset of thermal pressurization is also important in earthquake science to explain dynamic fault weakening during coseismic slip (Rempel and Rice 2006, Sulem et al. 2007). Important theoretical advances have been proposed in the study of thermal weakening of fault during coseismic slip and one can find an extensive literature review on the subject in the comprehensive paper of Rice (2006). In this paper, Rice emphasises the need of laboratory data to constrain theoretical modelling of these mechanisms. In particular thermal pressurization of rocks during seismic slip is highly influenced by damage and inelastic deformation inside the fault zones. The presence of clay material in fault zones also affects thermal pressurization as possible collapse of the clay under thermal loading may activate fluid pressurization (Sulem et al. 2007).

The values of the undrained thermal pressurization coefficient, defined as the pore pressure increase due to a unit temperature increase in undrained condition, is thus largely dependent upon the nature of the material, the state of stress, the range of temperature change, the induced damage. In the literature we can find values that differ from two orders of magnitude: In Campanella and Mitchell (1968) different values are found from 0.01 MPa/°C for clay to 0.05 MPa/°C for sandstone. Palciauskas and Domenica (1982) estimate a value of 0.59MPa/°C for Kayenta sandstone. On the basis of Sultan (1997) experimental data on Boom clay, Vardoulakis (2002) estimates this coefficient as 0.06 MPa/°C. For a clayey fault gouge extracted at a depth of 760m in Aigion fault in the Gulf of Corinth (Greece), the value obtained by Sulem et al. (2004) is 0.1 MPa/°C and for intact rock at great depth, the value given by Lachenbruch (1980) is 1.5 MPa/°C. For a mature fault at 7km depth at normal stress of 196 MPa, ambient pore pressure of 70 MPa, and ambient temperature of 210°C, Rice (2006) estimates this coefficient as 0.92MPa/°C in case of intact fault walls and 0.31MPa/°C in case of damage fault wall. The thermal pressurization coefficient of a hardened cement paste is evaluated experimentally by Ghabezloo *et al*. (2009), equal to 0.6 MPa/°C.

The large variability of the thermal pressurization coefficient in different geomaterials highlights the necessity of laboratory studies. The aim of this paper is to present the theoretical basis of thermal pressurization phenomenon and some aspects of its experimental evaluation.

## 2. Theoretical framework

The equations governing the phenomenon of the thermo-mechanical pressurization of porous materials are presented here for an ideal porous material characterized by a fully connected pore space and by a microscopically homogeneous (i.e., composed of only one solid material) and isotropic solid phase. These equations can be derived using the approach of Bishop and Eldin (1950). In this approach, schematized on Figure (1), the problem of thermo-mechanical loading is broken up into three independent sub-problems. Assuming isotropic elasticity, the volumetric changes of the porous material element and its components, are written separately, as presented on table (1). In this table, $V$ is the total volume and $n$ is the Lagrangian porosity defined as the pore volume per unit volume of porous material in the reference state. $\Delta\sigma$, $\Delta T$ and $\Delta u$ are the variations of the mean stress (positive in compression), the temperature and the pore pressure respectively. $c_d$ and $c_s$ are respectively the drained compressibility of the porous material and the compressibility of the solid phase. $\alpha_s$ is the volumetric thermal expansion coefficient of the solid phase and $c_f$ and $\alpha_f$ are respectively the compressibility and the thermal expansion coefficient of the pore fluid. The index 0 denotes the reference state.

The problem (a) in the Figure (1) corresponds to a particular loading case, called 'unjacketed' loading case, where an isotropic stress and a pore pressure of equal magnitudes are simultaneously applied to a volume element. This loading is as if the sample is submerged, without a jacket, in a fluid under pressure and results in a uniform pressure distribution in the solid phase. In the case of an ideal porous material, the material would deform as if all the pores were filled with the solid component and the skeleton and the solid component experience a uniform volumetric strain (Detournay and Cheng, 1993).

$$\frac{\Delta V_s}{V_s} = \frac{\Delta V_n}{V_n} = \frac{\Delta V}{V} \tag{1}$$

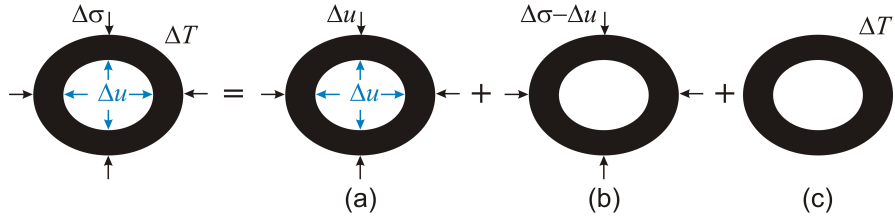

Figure 1: Decomposition of the problem of a thermo-mechanical loading into three independent problems

| Problem | Variation of volume of constituents | | Variation of volume of element |
|---|---|---|---|
| | Pore fluid | Solid phase | |
| (a) | $-n_0 V_0 c_f \Delta u$ | $-(1-n_0)V_0 c_s \Delta u$ | $-V_0 c_s \Delta u$ |
| (b) | - | $-V_0 c_s (\Delta\sigma - \Delta u)$ | $-V_0 c_d (\Delta\sigma - \Delta u)$ |
| (c) | $n_0 V_0 \alpha_f \Delta T$ | $(1-n_0)V_0 \alpha_s \Delta T$ | $V_0 \alpha_s \Delta T$ |

Table 1: Variation of volume of the porous material and its constituents for the three independent problems presented in Figure (1)

The problem (b) in the Figure (1) corresponds to Terzaghi effective stress loading. In this case the isotropic stress acting on the solid phase is $(\Delta\sigma - \Delta u)/(1-n_0)$ and the corresponding variation of volume is given by:

$$\Delta V_s = -(1-n_0)V_0 c_s \frac{\Delta\sigma - \Delta u}{1-n_0} = -V_0 c_s (\Delta\sigma - \Delta u) \tag{2}$$

The problem (c) in the Figure (1) is a drained thermal loading under constant total stress. For an ideal porous material, the variations of volume of the porous element and also one of the pore volume are characterized by the thermal expansion coefficient of the solid phase $\alpha_s$, because an isotropic thermal expansion would cause a proportional change in every linear dimension of the body (McTigue 1986). Assuming incremental linear thermo-elasticity, the volumetric strain of the element $\varepsilon_v$ (positive in contraction) is obtained as the sum of the variations of volume in the three sub-problems per unit total volume (in reference state):

$$\varepsilon_v = -\frac{\Delta V}{V_0} = c_s \Delta u + c_d (\Delta\sigma - \Delta u) - \alpha_s \Delta T \tag{3}$$

The variation of the porosity $n$, can be written as the difference between the volume change of the element and the volume change of the solid phase:

$$\Delta n = \frac{\Delta V - \Delta V_s}{V_0} = \left[ -c_s \Delta u - c_d (\Delta \sigma - \Delta u) + \alpha_s \Delta T \right]$$
$$- \left[ -(1-n_0) c_s \Delta u - c_s (\Delta \sigma - \Delta u) + (1-n_0) \alpha_s \Delta T \right]$$
(4)

Equation (4) can be simplified to obtain the following expression for the elastic change of porosity in drained condition:

$$\Delta n = -(c_d - c_s) \Delta \sigma + n_0 c_n \Delta u + n_0 \alpha_n \Delta T \tag{5}$$

where $c_n$ and $\alpha_n$ are respectively the compressibility and the volumetric thermal expansion coefficient of pore-volume defined by the following expressions:

$$c_n = \frac{1}{n_0} \left[ c_d - (1+n_0) c_s \right] \tag{6}$$

$$\alpha_n = \alpha_s \tag{7}$$

The undrained condition is defined theoretically as a condition in which there is no change in the pore fluid mass of the material. For a saturated porous material the fluid mass variation is written as $\Delta m_f = V_n \Delta \rho_f + \rho_f \Delta V_n$. By setting $\Delta m_f = 0$ and from the definitions of the fluid compressibility and thermal expansion coefficient one obtains:

$$\frac{\Delta V_n}{V_n} = -\frac{\Delta \rho_f}{\rho_f} = -c_f \Delta u + \alpha_f \Delta T \tag{8}$$

From the definition of the porosity and by using only the first order terms we have:

$$\frac{\Delta V_n}{V_n} \simeq \frac{\Delta n}{n_0} \tag{9}$$

Replacing equations (8) and (9) in equation (5) the following expression is obtained for the pore pressure change due to an undrained thermo-mechanical loading.

$$\Delta u = B \Delta \sigma + \Lambda \Delta T \tag{10}$$

where $B$ is the Skempton's (1954) coefficient:

$$B = \frac{c_d - c_s}{n_0 (c_f + c_n)} \tag{11}$$

and $\Lambda$ is the coefficient of thermal pressurization, as presented by Rice (2006):

$$\Lambda = \frac{\alpha_f - \alpha_n}{c_f + c_n} \tag{12}$$

Equation (12) clearly highlights that the discrepancy between the thermal expansion of the pore fluid and that of the pore volume is the factor causing the thermal pressurization of porous materials.

By replacing $\Delta u$ in equation (3) with the expression given in equation (10), one finds the following expression for the volumetric deformation of a porous material subjected to a thermo-mechanical loading in undrained condition:

$$\varepsilon_v = c_u \Delta\sigma - \alpha_u \Delta T \tag{13}$$

where $c_u$ and $\alpha_u$ are respectively the undrained compressibility and the undrained volumetric thermal expansion coefficient of porous material expressed by the following equations:

$$c_u = c_d - B(c_d - c_s) \tag{14}$$

$$\alpha_u = \alpha_s + \Lambda(c_d - c_s) \tag{15}$$

Using the equations (11) and (12) the expression of the undrained thermal expansion coefficient of a thermo-elastic porous material given by McTigue (1986) is retrieved.

$$\alpha_u = \alpha_s + Bn_0(\alpha_f - \alpha_s) \tag{16}$$

## 2.1 Effect of non-elastic strains

The above framework can be extended to account for the effect of non-elastic strains. These strains can be plastic, viscoelastic or viscoplastic and induce non-elastic porosity changes. The non-elastic changes of the total volume, pore volume and solid volume are defined by:

$$\Delta V^{ne} = \Delta V - \Delta V^e \; ; \; \Delta V_n^{ne} = \Delta V_n - \Delta V_n^e \; ; \; \Delta V_s^{ne} = \Delta V_s - \Delta V_s^e \tag{17}$$

The equation (3) can be re-written with the additional contribution of the non-elastic volume changes:

$$\varepsilon_v = -\frac{\Delta V}{V_0} = c_s \Delta u + c_d(\Delta\sigma - \Delta u) - \alpha_s \Delta T + \varepsilon_v^{ne} \tag{18}$$

In the same way, equation (5) can be re-written taking into account the non-elastic variation of the porosity $\Delta n^{ne}$:

$$\Delta n = -(c_d - c_s)\Delta\sigma + n_0 c_n \Delta u + n_0 \alpha_n \Delta T + \Delta n^{ne} \tag{19}$$

The non-elastic variation of porosity $\Delta n^{ne}$ can be calculated from the definition of the porosity and knowing that $V_n = V - V_s$.

$$\Delta n^{ne} = \frac{\Delta V_n^{ne}}{V_0} = -\varepsilon_v^{ne} + (1-n_0)\varepsilon_s^{ne} \tag{20}$$

Using equations (8), (9) and (20) in equation (19) the following relation is obtained for the variations pore pressure with the confining pressure and temperature in undrained condition, in presence of non-elastic volume changes:

$$\Delta u = B\Delta\sigma + \Lambda\Delta T + \frac{\varepsilon_v^{ne} - (1-n_0)\varepsilon_s^{ne}}{n_0(c_f + c_n)} \quad (21)$$

Equation (21) shows that non-elastic volume changes add an additional term in the generated pore pressure. The contracting volume changes, like in the case of thermal collapse of normally consolidated clays, will induce a pore pressure increase, while the dilating volume changes, as for example in the case of damage of fault walls, will induce a pore pressure reduction. In the case of a material for which the solid phase is elastic ($\varepsilon_s^{ne} = 0$), equation (21) can be rewritten as:

$$\Delta u = B\Delta\sigma + \Lambda\Delta T + \frac{\varepsilon_v^{ne}}{n_0(c_f + c_n)} \quad (22)$$

## 3. Evaluation of thermal pressurization coefficient

The thermal pressurization coefficient can be evaluated indirectly using equation (12) by knowing the thermoelastic properties of the material and the physical properties of the pore fluid. The thermal expansion coefficient and the compressibility of water are known as functions of temperature and pore pressure (Spang, 2002). Figure (2) present the variations of the thermal expansion coefficient and the compressibility of water for temperature between 10°C and 90°C and pore pressures between 1 MPa and 60 MPa (Spang, 2002). One can see that the variation of the thermal expansion coefficient with temperature is very important and have a significant effect on the evaluation of the thermal pressurization coefficient.

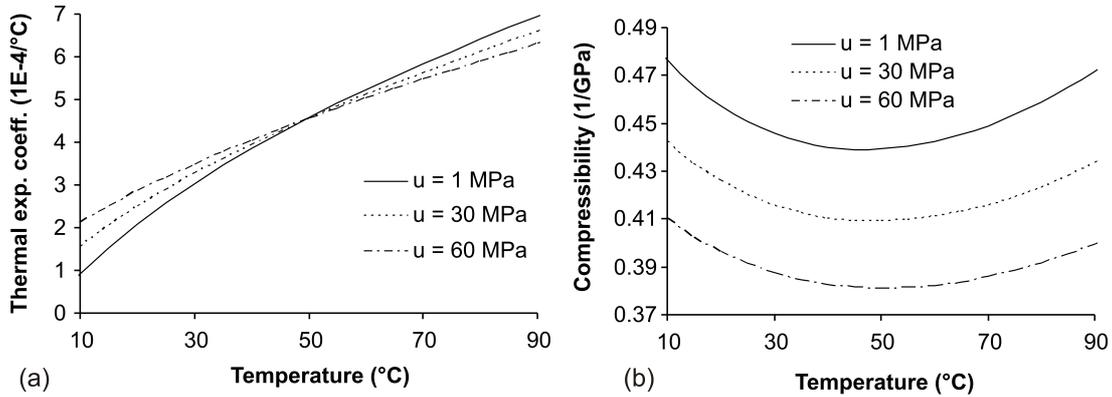

**Figure 2: Variation of the physical properties of water with temperature and pressure (a) thermal expansion coefficient, (b) compressibility**

For evaluation of the thermal pressurization coefficient, the pore volume compressibility $c_n$ should be evaluated using equation (6). The drained compressibility of the material should be known from the results of drained compression tests. In the case of a simple porous material which is constituted of one single solid component, the compressibility $c_s$ is equal to the compressibility of the solid component. Table 2 presents the bulk moduli of some commonly observed minerals. In the case of a porous material which is constituted of several solid components, the unjacketed compressibility $c_s$ can be estimated by knowing the mineralogy of the rock and the compressibility of each constituent. The homogenized unjacketed modulus can be evaluated using Hill's (1952) average formula which is simply the mid-sum of the Voigt (upper) and Reuss (lower) bounds:

$$K_s^{\text{hom}} = \frac{1}{2}\left[\sum f_i K_s^{(i)} + \left(\sum \frac{f_i}{K_s^{(i)}}\right)^{-1}\right] \tag{23}$$

where $f_i$ and $K_s^{(i)}$ are the volume fraction and the compression modulus of the $i^{\text{th}}$ constituent respectively.

| Mineral | Quartz | Calcite | Dolomite | Feldspars | Clay minerals |
|---|---|---|---|---|---|
| Bulk modulus (GPa) | 38[1] | 73[1] | 95[1] | 69[1] | 50[3] |
| Volumetric thermal expansion (×10⁻⁶ (°C)⁻¹) | 24.3[2] 33.4[4] | 3.8[2] | 22.8[2] | 8.9-15.6[2] | 34[3] |
| [1] Bass (1995), [2] Fei (1995), [3] McTigue (1986), [4] Palciauskas and Domenico (1982) | | | | | |

**Table 2: Bulk modulus and volumetric thermal expansion coefficient of some minerals**

When the porous material is constituted of a single solid component, the pore volume thermal expansion coefficient $\alpha_n$ is equal to the thermal expansion coefficient of the solid component. Some typical values of volumetric thermal expansion coefficient of minerals are presented in Table (2). For a porous material which is constituted of several solid components, the homogenized coefficient of thermal expansion of the solid phase $\alpha_s$ can also be evaluated using homogenisation theory, knowing the mineralogy of the rock and the thermal expansion coefficients of its constituents. The following equation can be used to calculate the average thermal expansion coefficient of a two phase linear thermo-elastic composite material (Berryman 1995, Zaoui 2000):

$$\alpha_s^{\text{hom}} = \left(f_1 \alpha_s^{(1)} + f_2 \alpha_s^{(2)}\right) + \frac{\frac{1}{K_s^{\text{hom}}} - \left(\frac{f_1}{K_s^{(1)}} + \frac{f_2}{K_s^{(2)}}\right)}{\frac{1}{K_s^{(2)}} - \frac{1}{K_s^{(1)}}} \left(\alpha_s^{(2)} - \alpha_s^{(1)}\right) \tag{24}$$

where $\alpha_s^{(i)}$ is the thermal expansion coefficient of the $i^{\text{th}}$ constituent. The other parameters are introduced before in equation (23). A uniform temperature variation imposed on a heterogeneous material induces residual thermal stresses due to thermal expansion mismatch between the different phases. This effect is reflected in the second term of the left-hand side of equation (24) as the compressibility of the constituents and of the composite material affects its thermal expansion.

# 4. Laboratory testing

The experimental evaluation of thermal pressurization coefficient can be performed in an undrained heating test. In a triaxial cell, this test is performed on a fully saturated sample under a constant confining pressure. During the test the temperature is varied at a constant rate or in several steps and the resulting pore pressure change is recorded. The effective stress is thus changing during the test. If the compressibility of the porous material is stress-dependent, the evaluated thermal pressurization coefficient will be influenced by changes of effective stress.

The undrained condition is defined theoretically as a condition in which there is no change in the fluid mass of the porous material. For performing an undrained test in the laboratory, this condition cannot be achieved just by closing the valves of the drainage system as it is done classically in a conventional triaxial system (Figure (3)). In a triaxial cell, the tested sample is connected to the drainage system of the cell and also to the pore pressure transducer. As the drainage system has a non-zero volume filled with fluid, it experiences volume changes due to its compressibility and its thermal expansivity. The variations of the volume of the drainage system and of the fluid filling the drainage system induce a fluid flow into or out of the sample to achieve pressure equilibrium between the sample and the drainage system. This fluid mass exchanged between the sample and the drainage system modifies the measured pore pressure and consequently the measured strains during the test.

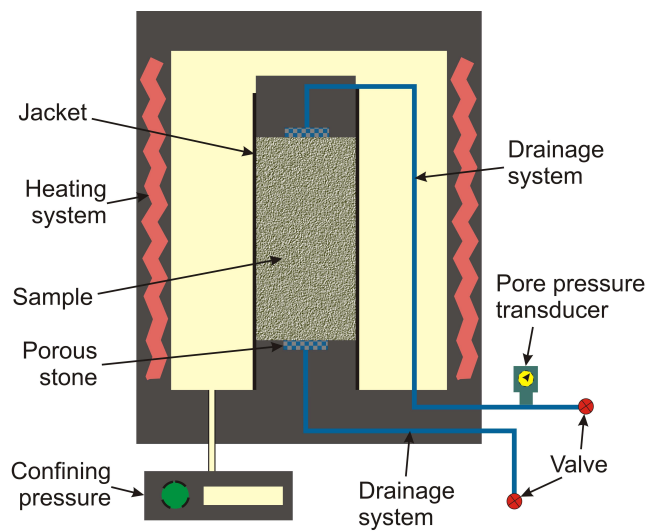

**Figure 3- Schematic view of a conventional triaxial cell**

Wissa (1969) was the first who studied this problem for a mechanical undrained loading. He presented an expression for the measured pore pressure increase as a function of the compressibilities of pore-water, soil skeleton, pore-water lines and pressure measurement system, but the compressibility of the solid phase was not considered in his work. Bishop (1976) presented an extension of the work of Wissa (1969) taking into account the compressibility of the solid grains. The proposed method was first used by Mesri *et al*. (1976) to correct the measured pore pressure in undrained isotropic compression tests. Ghabezloo and Sulem (2009a) have presented an extension to the work of Bishop (1976) to correct the pore pressure measured during undrained heating and cooling tests by taking into account the thermal expansion of the drainage system, the inhomogeneous temperature distribution in the drainage system and also the thermal expansion of the fluid filling the drainage system. The proposed method was applied to the results of undrained heating tests performed on Rothbach sandstone (Ghabezloo and Sulem, 2009a) and on a hardened cement paste (Ghabezloo *et al*., 2009). The correction method was then extended for the correction of the measured strains and consequently the correction of the measured undrained thermal expansion coefficient in Ghabezloo and Sulem (2009b). This correction method is summarized in the following section.

## *4.1 Correction of the effect of drainage system*

In a triaxial cell the tested sample is connected to the drainage system of the cell and the undrained condition is achieved by closing the valves of this system (Figure (3)). Consequently, the condition $\Delta m_f = 0$ is applied to the total volume of the fluid which fills the pore volume of the sample and also the drainage system:

$$m_f = V_n \rho_f + V_L \rho_{fL} \tag{25}$$

where $V_L$ is the volume of the drainage system and $\rho_{fL}$ is the density of the fluid in the drainage system. As the sample and the drainage system may have different temperatures and considering that the fluid density varies with temperature, different densities are considered for the pore-fluid of the sample and for the fluid filling the drainage system. The variation of volume of the drainage system can be written in the following form:

$$\frac{\Delta V_L}{V_L} = c_L \Delta u + \alpha_L \Delta T_L - \kappa_L \Delta \sigma \tag{26}$$

where $\Delta T_L$ is the equivalent temperature change in the drainage system, $c_L$ and $\kappa_L$ are isothermal compressibilities and $\alpha_L$ is the thermal expansion coefficient of the drainage system defined as:

$$c_L = \frac{1}{V_L}\left(\frac{\partial V_L}{\partial u}\right)_{T_L,\sigma} \tag{27}$$

$$\alpha_L = \frac{1}{V_L}\left(\frac{\partial V_L}{\partial T_L}\right)_{u,\sigma} \tag{28}$$

$$\kappa_L = -\frac{1}{V_L}\left(\frac{\partial V_L}{\partial \sigma}\right)_{u,T_L} \tag{29}$$

The parameter $c_L$ is equivalent to $(c_L + c_M)/V_L$ in Wissa (1969) and Bishop (1976). Based on the data provided by these authors, some typical values of $c_L$ can be evaluated which vary between 0.008 GPa⁻ [1] and 0.16 GPa$^{-1}$. It should be mentioned that the parameters $c_L$ and $\alpha_L$ defined in equations (27) and (28) are equivalent respectively to $c_L/V_L$ and $\alpha_L/V_L$ in Ghabezloo and Sulem (2009a).

In most triaxial devices, the drainage system can be separated into two parts: one situated inside the triaxial cell and the other one situated outside the cell. In the part inside the cell, one can assume that the temperature change $\Delta T$ is identical to the one of the sample; in the other part situated outside the cell, the temperature change is smaller than $\Delta T$ and varies along the drainage lines. We define an equivalent homogeneous temperature change $\Delta T_L$ such that the volume change of the entire drainage system caused by $\Delta T_L$ is equal to the volume change induced by the true non-homogeneous temperature field. The temperature ratio $\beta$ is an additional parameter which is defined below and evaluated on a calibration test as explained further:

$$\beta = \frac{\Delta T_L}{\Delta T} \tag{30}$$

By writing the undrained condition $\Delta m_f = 0$, using equation (26) and taking into account the variations of the fluid density with pore pressure and temperature changes, the following expression is obtained for the variations of the pore volume of the tested sample:

$$\frac{\Delta V_n}{V_n} = -c_f \Delta u + \alpha_f \Delta T + \frac{V_L}{V_n}\frac{\rho_{fL}}{\rho_f}\left(-c_{fL}\Delta u - c_L \Delta u + \alpha_{fL}\beta\Delta T - \alpha_L \beta \Delta T + \kappa_L \Delta \sigma \right) \tag{31}$$

As the thermal expansion coefficient and the compressibility of the pore fluid both vary with temperature, the parameters used for the fluid in the drainage system, $\alpha_{fL}$ and $c_{fL}$, are different from the parameters used for the pore fluid of the porous material. Replacing equations (31) and (9) in equation (5) the expressions of the measured Skempton coefficient and thermal pressurization coefficient, $B^{mes}$ and $\Lambda^{mes}$ are obtained.

$$B^{mes} = \frac{(c_d - c_s) + \dfrac{V_L}{V_0}\dfrac{\rho_{fL}}{\rho_f}\kappa_L}{n_0(c_f + c_n) + \dfrac{V_L}{V_0}\dfrac{\rho_{fL}}{\rho_f}(c_{fL} + c_L)} \tag{32}$$

$$\Lambda^{mes} = \frac{n_0(\alpha_f - \alpha_n) + \beta\dfrac{V_L}{V_0}\dfrac{\rho_{fL}}{\rho_f}(\alpha_{fL} - \alpha_L)}{n_0(c_f + c_n) + \dfrac{V_L}{V_0}\dfrac{\rho_{fL}}{\rho_f}(c_{fL} + c_L)} \tag{33}$$

The comparison of the equations (32) and (33) respectively with equations (11) and (12) shows the effect of the drainage system of the triaxial cell on the measured coefficients. Using equations (32), (11) and (14) the expressions of the corrected Skempton coefficient $B^{cor}$ and the corrected undrained bulk compressibility $c_u^{cor}$ are found:

$$B^{cor} = \frac{B^{mes}}{1 + \dfrac{V_L \rho_{fL}}{V_0 \rho_f (c_d - c_s)} \left[ \kappa_L - B^{mes} (c_{fL} + c_L) \right]} \quad (34)$$

$$c_u^{cor} = c_d - \frac{c_d - c_u^{mes}}{1 + \dfrac{V_L \rho_{fL}}{V_0 \rho_f (c_d - c_s)} \left[ \kappa_L - \dfrac{c_d - c_u^{mes}}{c_d - c_s} (c_{fL} + c_L) \right]} \quad (35)$$

Similarly using equations (33), (12) and (15) the following expressions are obtained for the corrected thermal pressurization coefficient $\Lambda^{cor}$ and the corrected undrained thermal expansion coefficient $\alpha_u^{cor}$:

$$\Lambda^{cor} = \frac{\Lambda^{mes}}{1 + \dfrac{V_L \rho_{fL}}{n_0 V_0 \rho_f (\alpha_f - \alpha_n)} \left[ \beta(\alpha_{fL} - \alpha_L) - \Lambda^{mes} (c_{fL} + c_L) \right]} \quad (36)$$

$$\alpha_u^{cor} = \alpha_d + \frac{\alpha_u^{mes} - \alpha_d}{1 + \dfrac{V_L \rho_{fL}}{n_0 V_0 \rho_f (\alpha_f - \alpha_n)} \left[ \beta(\alpha_{fL} - \alpha_L) - (\alpha_u^{mes} - \alpha_d) \dfrac{c_{fL} + c_L}{c_d - c_s} \right]} \quad (37)$$

Equation (34) is similar to the one presented by Bishop (1976), but differs in two points. The first one is that Bishop did not account for the influence of the confining pressure on the dead volume of the drainage system. This effect appears in equation (34) through the parameter $\kappa_L$. The second one is that Bishop assumed equal densities and compressibilities for the sample's pore fluid and the fluid in the drainage system, which is a correct assumption for an isothermal undrained test performed at ambient temperature. The same assumption is also made by Ghabezloo and Sulem (2009a). For an isothermal undrained test performed at an elevated temperature, the situation depends on the heating system of the triaxial cell. If the heating system is such that the temperature change is uniform in the sample and in the drainage system, the assumption of similar properties for the sample's pore fluid and the fluid in the drainage system can be used. Otherwise, as shown in Figure (3), the sample temperature is different from the average temperature in the drainage system so that different properties should be considered for the sample's pore fluid and the fluid in the drainage system.

The correction method proposed here in equations (34) and (35) is applied directly on the results of the test, but it is restricted to an elastic response of the sample and of the drainage system. It differs from the method proposed by Lockner and Stanchits (2002) who have modified the procedure of the test itself by imposing a computer-generated virtual 'no-flow boundary condition' at the sample-endplug interface to insure that no volume change occurs in the drainage system.

## *4.2 Calibration of the correction parameters*

The drainage system is composed of all the parts of the system which are connected to the pore volume of the sample and filled with the fluid, including pipes, pore pressure transducers, porous stones. The volume of fluid in the drainage system $V_L$, can be measured directly or evaluated by using the geometrical dimensions of the drainage system. For the triaxial cell used in the present study, the volume of the drainage system was measured directly using a pressure/volume controller. As can be seen in Figure (3), the drainage system has two main parts: one connected to the top and the other connected to the bottom of the sample. Before performing the measurement, the drainage system was emptied using compressed air and then each part was connected to the pressure/volume controller keeping the valve of the drainage system closed. The connection pipe between the pressure/volume controller and the drainage system was filled with water. Then a small pressure was applied by the pressure/volume controller and its volume was set to zero before opening the valve of the drainage system. By opening the valve, the fluid flows into the drainage system. As soon as the first drop of the fluid flows out of the porous stone, the pressure/volume controller is stopped and the volume of the fluid which has filled the corresponding part of the drainage system is directly given by the volume change of the pressure/volume controller. The measurement was repeated for each part of the drainage system and the total volume of the drainage system was evaluated equal to $V_L = 2300\,\text{mm}^3$.

The compressibility of the drainage and pressure measurement systems $c_L$ is evaluated by applying a fluid pressure and by measuring the corresponding volume change in the pressure/volume controller. A metallic sample is installed inside the cell to prevent the fluid to go out from the drainage system. Fluid mass conservation is written in the following equation which is used to calculate the compressibility $c_L$ of the drainage system:

$$\frac{\Delta V_L}{V_L} = \left(c_L + c_{fL}\right)\Delta u \tag{38}$$

$\Delta u$ and $\Delta V_L$ are respectively the applied pore pressure and the volume change measured by the pressure/volume controller. For a single measurement, the volume change $\Delta V_L$ accounts also for the compressibility of the pressure/volume controller and of the lines used to connect the pressure/volume controller to the main drainage system. To exclude the compressibility of these parts, a second measurement is done only on the pressure/volume controller and the connecting lines. The volume change $\Delta V_L$ used in equation (38) is the difference between these two measurements. The measurements were performed separately for the two parts of the drainage system with volume $V_{L1}$ and $V_{L2}$ respectively. The compressibility of the entire system $c_L$ is simply obtained as the weighted average of the compressibilities of each part ($c_{L1}$ and $c_{L2}$ respectively):

$$c_L = \frac{V_{L1}}{V_L} c_{L1} + \frac{V_{L2}}{V_L} c_{L2} \tag{39}$$

The estimated value is $c_L = 0.117 \text{GPa}^{-1}$. This is equivalent to the compressibility that can be obtained in a single measurement in which the pressure/volume controller is connected simultaneously to the both parts of the drainage system using a T-connection.

The parameter $\beta$ and the thermal expansivity of the drainage system $\alpha_L$ are evaluated using the results of an undrained heating test performed using a metallic sample with the measurement of the fluid pressure change in the drainage system. For the metallic sample $n_0 = 0$ and $c_d = c_s$ so that equation (33) is reduced to the following equation:

$$\Lambda^{mes} = \frac{\beta(\alpha_{fL} - \alpha_L)}{c_{fL} + c_L} \tag{40}$$

The thermal expansion coefficient $\alpha_{fL}$ and the compressibility $c_{fL}$ of water are known as functions of temperature and fluid pressure. As these variations are highly non-linear, the parameters $\beta$ and $\alpha_L$ cannot be evaluated directly but are back analysed from the calibration test results: the undrained heating test of the metallic sample is simulated analytically using equation (40) with a step by step increase of the temperature. For each step the corresponding water thermal expansion and compressibility are used (Spang 2002). The parameters $\beta$ and $\alpha_L$ are back-calculated by minimizing the error between the measurements and the computed results using a least-square algorithm. The test result and the back analysis are presented in Figure (4). The parameter $\beta$ is found equal to 0.46 and the thermal expansion coefficient of the drainage system $\alpha_L$ for this test is found equal to $1.57 \times 10^{-4}$ (°C)$^{-1}$.

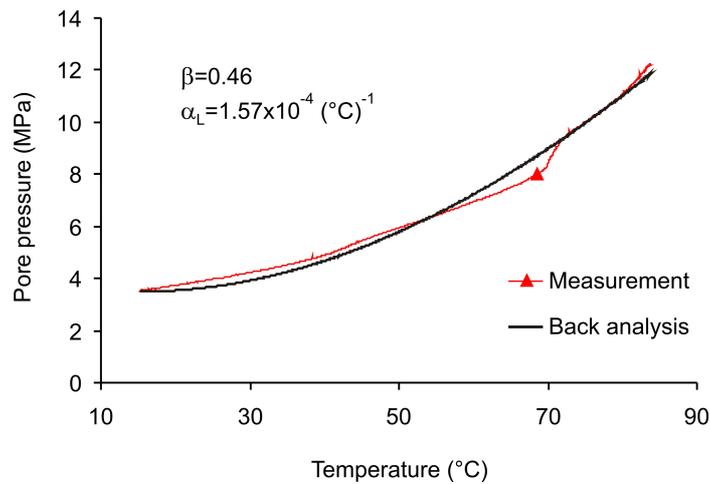

**Figure 4- Calibration test for the evaluation of the temperature ratio $\beta$ and the thermal expansion $\alpha_L$ of the drainage system-comparison.**

The evaluation of the compressibility $\kappa_L$ which represents the effect of the confining pressure on the volume of the drainage system is performed using an analytical method. As can be seen in Figure (3), only a part of the drainage system which is the pipe connected to the top of the sample, is influenced by the confining pressure. The effect of the confining pressure on the variations of the volume of this pipe can be evaluated using the elastic solution of the radial displacement of a hollow cylinder. Considering a hollow cylinder with inner radius $a$ and outer radius $b$ and radial stresses $p_i$ and $p_o$, respectively at the inner and outer boundaries, the following expression is obtained for the compressibility $\kappa_L$ using the well-known Lamé solution:

$$\kappa_L = \frac{4\pi a^2 b^2 L(1-v^2)}{(b^2-a^2)V_L E} \tag{41}$$

where $E$ and $v$ are respectively Young's modulus and Poisson's coefficient and $L$ is the length of the pipe. The details of the derivation are presented in Ghabezloo and Sulem (2009b). For the considered pipe $a=0.25\text{mm}$, $b=0.8\text{mm}$, $L=900\text{mm}$, $E=190\text{GPa}$ and $v=0.3$. Inserting these values and $V_L=2300\text{mm}^3$ in equation (41) we obtain $\kappa_L=1.6\times10^{-3}\text{GPa}^{-1}$. This value is very small as compared to the compressibility $c_L=0.117\text{GPa}^{-1}$ which takes into account the effect of the pore pressure variations on the volume of the drainage system. This is due to the fact that only a small part of the drainage system, less than 8% of its volume, is influenced by the confining pressure.

## 4.3 Parametric study of measurement error

In this section, a parametric study on the error made on the measurement of the thermal pressurization coefficient and the undrained thermal expansion coefficient is presented. A similar parametric study for Skempton coefficient and the undrained compressibility is presented in Ghabezloo and Sulem (2009b). The error on a quantity $Q$ is evaluated as $(Q_{\text{measured}}-Q_{\text{real}})/Q_{\text{real}}$ and takes positive or negative values with indicates if the measurement overestimates or underestimates the considered quantity. As shown in the following, among the different parameters appearing in equations (36) and (37), the porosity $n_0$ of the tested material, its drained compressibility $c_d$ and the ratio of the volume of the drainage system to the one of the tested sample, $V_L/V_0$ are the most influent parameters. For this parametric study the parameters of the drainage system are taken equal to the ones of the triaxial system used in this study, as presented in section 4.2. We take also $c_s=0.02\text{GPa}^{-1}$ and $\alpha_n=3\times10^{-4}(°C)^{-1}$ which are typical values for minerals. Figure (5) present the error on the measurements of the thermal pressurization coefficient $\Lambda$ and of the undrained thermal expansion coefficient $\alpha_u$ as a function of the sample porosity $n_0$, for three different values of drained compressibility and two different values of the ratio $V_L/V_0$. Three different values of the drained compressibility are considered, respectively equal to 0.03GPa$^{-1}$, 0.1GPa$^{-1}$ and 0.5GPa$^{-1}$, which covers a range from a rock with a low compressibility to a relatively highly compressible rock. The porosity is varied from 0.05 to 0.35. The ratio $V_L/V_0$ is taken equal to 0.025 which corresponds to the conditions of the triaxial system used in this study. We analyze also the effect of a greater volume of the drainage system on the measurement errors by choosing another value twice bigger, equal to 0.05. We observe that the error of the measurement for $\Lambda$ varies between -40% and +10%, which shows that the measured value may be smaller or greater than the real one. Considering the tested material, the measurement error is more significant for low-porosity rocks with low-compressibility. We can also see the significant effect of the volume of the drainage system on the measurement error. The error for the undrained thermal expansion coefficient $\alpha_u$ varies between -6% and +4%, which is a narrower range, as compared to the thermal pressurization coefficient.

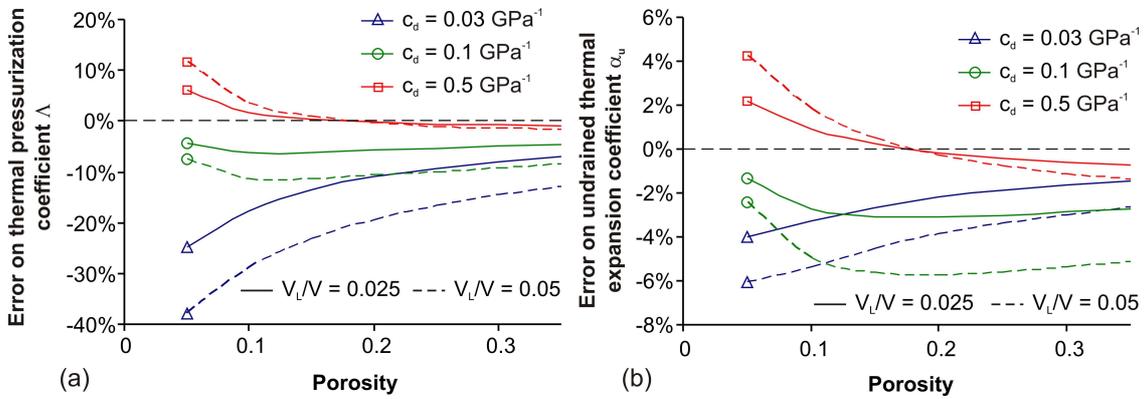

**Figure 5- Parametric study of the error on (a) the thermal pressurization coefficient $\Lambda$, (b) the undrained thermal expansion coefficient $\alpha_u$**

## 5. Example of an experimental study on a granular rock

An example of the experimental evaluation of undrained thermoelastic parameters and of the application of the correction method is presented for an undrained heating test performed on a fluid-saturated granular rock, Rothbach sandstone. The rock has a porosity of $n$=16% and is composed of 85% quartz, 12% feldspars and 3% clay. As can be seen in equations (36) and (37), the application of the correction method needs the knowledge of the drained and the unjacketed thermo-poro-elastic parameters of the tested material. The required parameters are evaluated and presented by Ghabezloo and Sulem (2009a). The laboratory experiments are performed on cylindrical sample with a diameter of 40mm and a height of 80mm. The volume of the sample is thus $V_0 = 100530 \text{mm}^3$ and the volume of the porous space is $nV_0 = 16085 \text{mm}^3$. The volume of the drainage system evaluated to 2300mm$^3$ is thus 14% of the volume of the porous space.

The unjacketed modulus $K_s = 1/c_s$ of the rock was evaluated in an unjacketed compression test and found equal to 41.6 GPa. A drained isotropic compression test was performed with a loading-unloading cycle and the tangent drained bulk modulus $K_d = 1/c_d$ of the rock was found to be stress-dependent and could be approximated by the following expression:

$$K_d = 0.96\,\sigma_d + 0.70 \qquad \sigma_d \leq 9\,\text{MPa}$$
$$K_d = 0.07\,\sigma_d + 8.72 \qquad \sigma_d > 9\,\text{MPa}$$

($K_d$ : GPa , $\sigma_d$ : MPa) (42)

## 5.1 Experimental setting

The triaxial cell used in this study can sustain a confining pressure up to 60MPa. It contains a system of hydraulic self-compensated piston. The loading piston is then equilibrated during the confining pressure build up and directly applies the deviatoric stress. The axial and radial strains are measured directly on the sample inside the cell with two axial transducers and four radial ones of LVDT type. The confining pressure is applied by a servo controlled high pressure generator. Hydraulic oil is used as confining fluid. The pore pressure is applied by another servo-controlled pressure generator with possible control of pore volume or pore pressure.

The heating system consists of a heating belt around the cell which can apply a temperature change with a given rate and regulate the temperature, and a thermocouple which measures the temperature of the heater. In order to limit the temperature loss, an insulation layer is inserted between the heater element and the external wall of the cell. A second insulation element is also installed beneath the cell. The heating system heats the confining oil and the sample is heated consequently. Therefore there is a discrepancy between the temperature of the heating element in the exterior part of the cell and that of the sample. In order to control the temperature in the centre of the cell, a second thermocouple is placed at the vicinity of sample. The temperature given by this transducer is considered as the sample temperature in the analysis of the test results. A schematic view of this triaxial cell is presented in Ghabezloo and Sulem (2009a) and Sulem and Ouffroukh (2006).

## 5.2 Drained heating test

In order to measure the drained thermal expansion coefficient of the rock, a drained heating test (see sub-problem (c) in Figure (1)) was carried out under a constant confining pressure. The initial temperature of the performed drained heating test was 21°C. The confining pressure and the pore water pressure during the test were maintained constant at 2.5 MPa and 1.0 MPa respectively. The drained thermal expansion coefficient is the slope of the temperature-volumetric strain response and is evaluated as $28 \times 10^{-6} \left(°\text{C}\right)^{-1}$ (Figure (6)). The coefficient of thermal expansion can also be evaluated indirectly using homogenisation theory, knowing the mineralogy of the rock and the thermal expansion coefficients of its constituents, as presented in equation (24). In order to evaluate the thermal expansion coefficient from this equation, we need first to evaluate the solid phase bulk modulus $K_s$ using equation (23). As mentioned above, the Rothbach sandstone contains 85% quartz, 12% feldspars and 3% clay. The values of elastic properties of different minerals are presented in Table (2). The compression modulus of quartz and the average compression modulus of feldspars are equal to 37.8 and 69.1 GPa respectively. The compression modulus of the clay solid grains is equal to 50 GPa. Using equation (23) the homogenized unjacketed modulus of Rothbach sandstone is evaluated equal to 41.1 GPa which is very close to the experimentally measured value, equal to 41.6 GPa. From Table (2) the thermal expansion coefficient of quartz is equal to $33.4 \times 10^{-6} (°C)^{-1}$ and the average thermal expansion coefficient of feldspars can be taken equal to $11.1 \times 10^{-6} (°C)^{-1}$. Considering the small volume fraction of clay in Rothbach sandstone, the equation (24) can be used to calculate the homogenized thermal expansion coefficient of solid grains by neglecting this part and taking into account only quartz and feldspar minerals. Using the given parameters and the $K_s^{hom}$ calculated with equation (23), the thermal expansion coefficient of Rothbach sandstone is found equal to $29.7 \times 10^{-6} (°C)^{-1}$. This value is in very good accordance with the measured thermal expansion coefficient equal to $28 \times 10^{-6} (°C)^{-1}$.

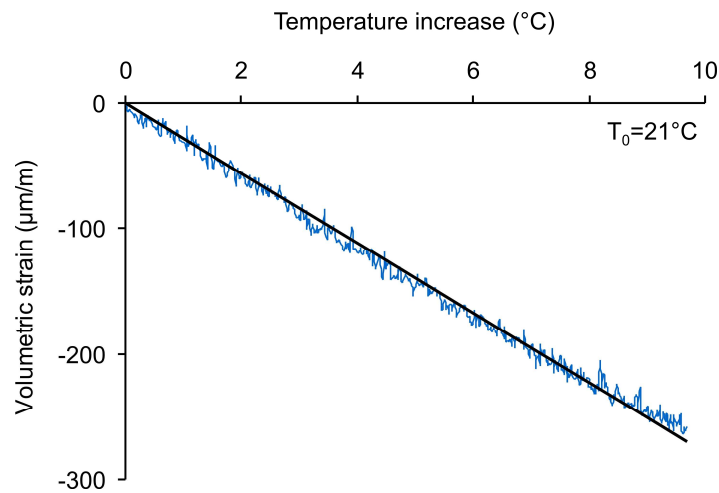

**Figure 6- Drained heating test on Rothbach sandstone: Volumetric strain response**

## 5.3 Undrained heating test

The undrained heating test was performed under constant isotropic stress equal to 10 MPa. The initial temperature of the sample was 20°C and the rate of temperature change was 0.2°C/min. The measured pore pressure during the test is presented in Figure (7a) as a function of the temperature increase. One can observe the non-linear increase of the pore pressure with the temperature up to the state for which the pore pressure in the sample reaches the confining pressure. At this point the pore fluid of the sample infiltrates between the sample and the rubber membrane so that the pore fluid pressurization is stopped. The slope of pore pressure curve versus the temperature gives the thermal pressurization coefficient $\Lambda$ and is presented in Figure (7b) as a function of the temperature increase. The non-linearity of the observed thermal pressurization coefficient is due to the (effective) stress-dependent compressibility of the sandstone and also to the temperature and pressure dependent compressibility and thermal expansion of the pore water. More details about the mechanism governing this non-linear behaviour can be found in Ghabezloo and Sulem (2009a). At the beginning of the test the volumetric strain could not be recorded due to a failure of the displacement sensors inside the triaxial cell. Therefore the volumetric strain-temperature curve, presented in Figure (8a), only starts from 40°C. The slope of this curve gives the undrained thermal expansion coefficient and is presented in Figure (8b) as a function of the temperature increase. We can see the increase of this coefficient with temperature which is mostly due to the significant increase of the thermal expansion coefficient of the water with temperature.

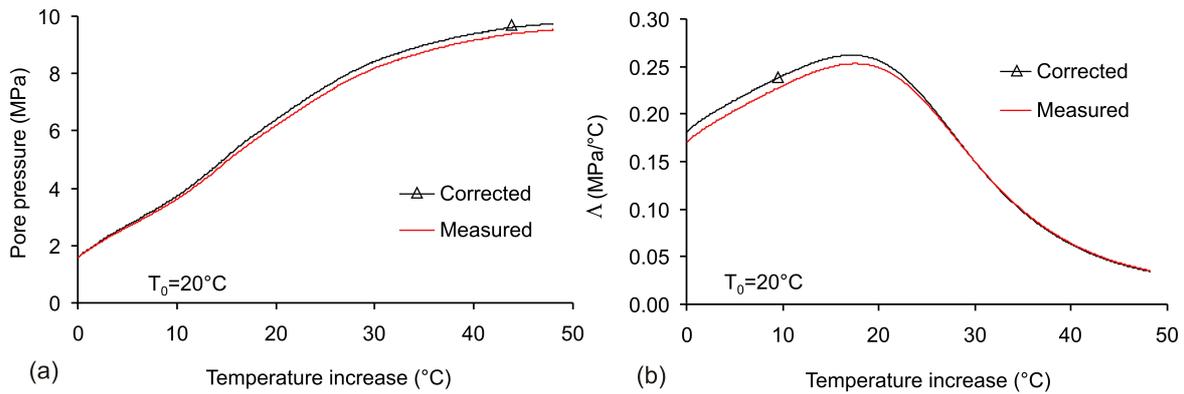

**Figure 7- Undrained heating test on Rothbach sandstone: (a) Measured and corrected pore pressure response, (b) Measured and corrected thermal pressurization coefficient**

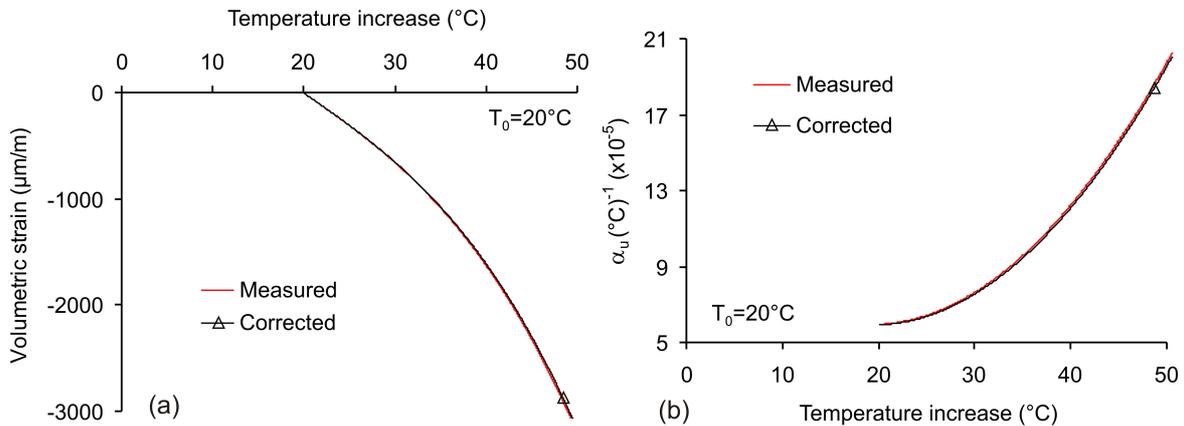

**Figure 8- Undrained heating test on Rothbach sandstone: (a) Measured and corrected volumetric strain response, (b) Measured and corrected undrained thermal expansion coefficient**

The test results are corrected using equations (36) and (37) for the effect of the dead volume of the drainage system. At each data point the relevant thermal expansion and compressibility of water are used as a function of the current pore pressure and temperature (Spang 2002). The drained compressibility is also calculated at each point as a function of Terzaghi effective stress (equation (42)). The variation of the mechanical properties of the sandstone with temperature is neglected in the analysis. It should be mentioned that the very good compatibility obtained by Ghabezloo and Sulem (2009a) between the results of an analytical simulation of this test and the experimental results showed the negligible effect of this assumption. The corrected values of the pore pressure, thermal pressurization coefficient, volumetric strain and undrained thermal expansion coefficient are presented along with the measured values respectively on Figures (7) and (8). The corrected pore pressure and thermal pressurization coefficient are bigger than the measured values. The corrected volumetric strain and undrained thermal expansion coefficient are slightly smaller than the measured values, but the average difference between the corrected and measured curves is about 1%. As can be seen in equations (36) and (37), apart from the properties of the drainage system, the correction depends also on the measured values of thermal pressurization coefficient and undrained thermal expansion coefficient. For the same triaxial cell, the correction of the observed pore pressure during an undrained heating test performed on a hardened cement paste (Ghabezloo *et al*. 2009) is more important than the one obtained here for Rothbach sandstone.

Using the non-linear elastic model presented in equation (42), the measured values of $\alpha_s$ and $c_s$ and knowing the compressibility and thermal expansion coefficient of water as functions of temperature and pore pressure, we can evaluate the thermal pressurization coefficient $\Lambda$ and the undrained thermal expansion coefficient of Rothbach sandstone as a function of temperature and effective stress, using equations (12) and (15). The calculated coefficients $\Lambda$ and $\alpha_u$ are presented on Figure (9) as a function of the effective stress up to 15 MPa and for different temperatures from 20°C to 90°C. The results show clearly the (effective) stress and temperature dependent character of the thermal pressurization coefficient and the undrained thermal expansion coefficient. The stress-dependency is significant in the range of stress-dependency of the rock compressibility (up to 9 MPa). The values of the coefficient $\Lambda$ vary from 0.02 to 0.72 MPa/°C depending on the effective stress and the temperature. The value of undrained thermal expansion coefficient $\alpha_u$ vary from $3\times10^{-5}\left(°C\right)^{-1}$ to $13\times10^{-5}\left(°C\right)^{-1}$ and is bigger for higher temperatures and smaller effective stresses. This strong non-linearity of the thermal response of the rock in undrained condition is mainly due to the variations of the thermal expansion and the compressibility of water with temperature and pore pressure. Moreover, the stress dependency of the compressibility of the material has important consequences on its thermal response in undrained condition.

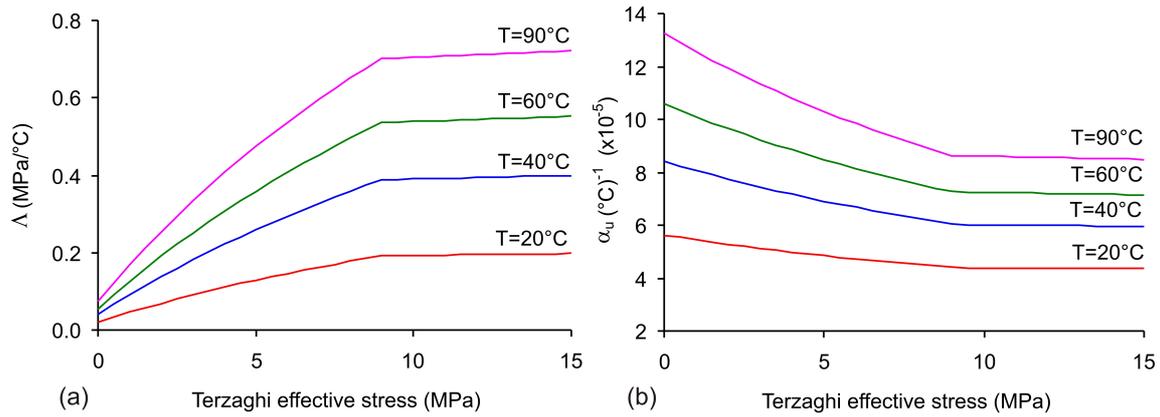

**Figure 9:** Effect of effective stress and temperature on (a) the thermal pressurization coefficient, (b) the undrained thermal expansion coefficient of Rothbach sandstone

# 6. Conclusion

The theoretical basis of the thermal response of the fluid-saturated porous materials in undrained condition is presented. The temperature increase in undrained condition leads to the pore pressure increase and deformation of the material. It has been demonstrated that the thermal pressurization phenomenon is controlled, on one hand by the discrepancy between the thermal behaviour of the pore fluid and of the solid phase, and on the other hand by the compressibility of the pore volume. The strong influence of temperature and stress on thermal pressurization of rocks is explained by the effect of temperature and pressure on the physical properties of water and by the stress-dependent character of the compressibility of porous rocks. Moreover, the important effect of the non-elastic strains of the material on the variations of the pore pressure has been also demonstrated. These non-elastic volume changes induce a pore pressure increase when they are contracting, while the dilating volume changes induce a pore pressure reduction.

For evaluation of the undrained thermo-poro-elastic properties of saturated porous materials in conventional triaxial cells, it is important to take into account the effect of the dead volume of the drainage system. The compressibility and the thermal expansion of the drainage system along with the dead volume of the fluid filling this system, influence the measured pore pressure and volumetric strain during an undrained thermal or mechanical loading in a triaxial cell. A simple correction method is presented to correct the measured pore pressure change and also the measured volumetric strain and consequently the evaluated thermal pressurization and undrained thermal expansion coefficients during an undrained heating test. A parametric study has demonstrated that the porosity of the tested material, its drained compressibility $c_d$ and the ratio of the volume of the drainage system to the one of the tested sample, $V_L/V_0$ are the key parameters which influence the most the error induced on the measurements by the drainage system. It has also shown that the measurement of thermal pressurization coefficient is much more affected than the measurement of the undrained thermal expansion coefficient.

An example of the experimental evaluation of undrained thermoelastic parameters and of the application of the correction method is presented for an undrained heating test performed on a fluid-saturated granular rock. The experimental results show the strong non-linearity of the thermal response of the rock in undrained condition. This non-linearity is mainly due to the variations of the thermal expansion and the compressibility of water with temperature and pore pressure. Moreover, the stress dependency of the compressibility of the material has important consequences on its thermal response in undrained condition.

# 7. Acknowledgments

The authors wish to thank François Martineau for his assistance in the experimental work.